\documentclass[prd,twocolumn,nofootinbib,reprint]{revtex4-1}
\usepackage{amsmath,amssymb,amsthm,bm,braket,ascmac,bbm,ulem}
\usepackage{graphicx}

\theoremstyle{definition}

\def\dd{\mathrm{d}}

\def\tot{{\rm tot}}

\def\Mpl{M_{\rm Pl}}
\def\GeV{{\rm GeV}}

\usepackage{color}

\begin{document}

\begin{flushleft}
RESCEU-2/20 
\end{flushleft}

\title{Gravitational Waves from Primordial Magnetic Fields\\[3pt]
via Photon-Graviton Conversion}
\author{Tomohiro Fujita$^1$, Kohei Kamada$^2$ and Yuichiro Nakai$^3$}
\affiliation{\vspace{2mm} $^1$Department of Physics, Kyoto University, Kyoto, 606-8502, Japan \\
$^2$Research Center for the Early Universe (RESCEU),
Graduate School of Science, The University of Tokyo, Tokyo 113-0033, Japan \\
$^3$Tsung-Dao Lee Institute and School of Physics and Astronomy, Shanghai
Jiao Tong University, 800 Dongchuan Road, Shanghai 200240, China}


\begin{abstract}
\vspace{3mm}

We explore a novel process in the early Universe in which  
thermalized photons are converted into gravitons in the presence of strong primordial magnetic fields.
It is found that the frequency of generated gravitational waves (GWs) is typically of the order of GHz, and their amplitude can be up to 
$ \Omega_\mathrm{GW}h^2 \sim 10^{-10}$.
If detected with future developments of the technology to explore this frequency region, the produced stochastic GW background 
enables us to know when and how strong the primordial magnetic fields are generated.
From the peak frequency of the GWs, we can also probe the number of relativistic degrees of freedom at that time.

\end{abstract}
\pacs{***}
\maketitle

\section{Introduction}\label{sec:intro}

The detection of gravitational waves (GWs) from binary compact star mergers 
such as those of black holes~\cite{Abbott:2016blz} and neutron stars~\cite{TheLIGOScientific:2017qsa} by the LIGO/Virgo collaboration
showed directly their very existence and opened the era of gravitational-wave astrophysics. 
We have now prospects for a wealth of quantitatively new astrophysical observations~\cite{Kuroda:2015owv}. 
Indeed, not only the ground-based interferometers LIGO and Virgo 
as well as the forthcoming KAGRA~\cite{Somiya:2011np} that have a sensitivity 
at $f \simeq 10^{1\sim3}$ Hz,  there are several on-going and proposed experiments 
at broad range of frequencies. 
The B-mode polarization of the cosmic microwave background (CMB) 
probes GWs with frequencies at $f\simeq 10^{-18\sim16}$ Hz~\cite{Zaldarriaga:1996xe,Kamionkowski:1996ks,Aghanim:2018eyx}. 
Pulsar timing arrays such as EPTA~\cite{Kramer:2013kea} and NANOGrav~\cite{McLaughlin:2013ira} observe GWs at 
$f \simeq 10^{-9\sim 7}$ Hz. 
The space-based interferometers  such as LISA~\cite{AmaroSeoane:2012km,Bartolo:2016ami}, DECIGO~\cite{Kawamura:2011zz}, and BBO~\cite{Crowder:2005nr} are sensitive 
at $f \simeq 10^{-3} \sim 10$ Hz. 
Moreover, recently new ideas to detect GWs with much higher frequencies
are also investigated~\cite{Cruise:2006zt,Akutsu:2008qv,Cruise:2012zz,Ito:2019wcb,Ejlli:2019bqj}. 
From the theoretical point of view, it is thus essential to extensively study 
various phenomena generating GWs with such a wide range of frequencies.

The stochastic GW background provides us with invaluable opportunities to access various processes in the early Universe, since the interaction of GWs with matter or radiation is so tiny that they directly carry information on their generation and global evolution history of the Universe~\cite{Maggiore:1999vm}. The stochastic GW background is produced by 
the quantum fluctuation during inflation~\cite{Starobinsky:1979ty}, 
particle production associated with inflation~\cite{Cook:2011hg},
violent processes at the end of inflation (i.e. preheating)~\cite{Khlebnikov:1997di}, cosmic strings~\cite{Vilenkin:1981bx}, as well as the first order phase transitions~\cite{Kosowsky:1992rz}. In addition to detailed investigation on the impact of the detection of GWs from these processes, it is important to look for other novel processes that produce stochastic GW backgrounds in the early Universe.

We here focus on the photon-graviton conversion process through strong magnetic fields~\cite{Gertsenshtein1962,DeLogi:1977qe,Zeldovich:1983cr,Cillis:1996qy}, 
which is similar to the photon-axion conversion process~\cite{Raffelt:1987im,Sikivie:1983ip,Csaki:2001yk,Deffayet:2001pc}. (See Ref.~\cite{Dolgov:2012be} for the opposite process.)
If electromagnetic waves (EMWs) are propagating in static magnetic fields, the energy momentum tensor has the quadrupole moment and GWs are produced. From the perspective of spins, a propagating electromagnetic
wave ({\it i.e.} a spin 1 field = photon) interacting with another spin 1 field, namely a magnetic field, can generate a spin 2 particle, graviton.

This conversion process has not yet been observed, because we cannot produce
sufficiently strong magnetic fields and EMWs that emit detectable GWs in the laboratory.
However, in the primordial Universe when the high energy photons exist, 
strong magnetic fields can also exist.\footnote{
Magnetogenesis scenarios from inflation~\cite{Turner:1987bw,Ratra:1991bn,Garretson:1992vt,Anber:2006xt,Fujita:2015iga,Fujita:2019pmi}, 
first order phase transitions~\cite{Vachaspati:1991nm,Baym:1995fk,Grasso:1997nx,Quashnock:1988vs,Cheng:1994yr,Sigl:1996dm}, 
or chiral plasma instability~\cite{Joyce:1997uy,Tashiro:2012mf,Brandenburg:2017rcb,Kamada:2018tcs}
have been considered.}
In fact, we have a strong motivation to consider their existence, 
since they can be the origin of the magnetic fields in the galaxies and galaxy clusters~\cite{Widrow:2002ud}. 
Moreover, recent observations (of the absence) of the inverse Compton cascade photon 
from the TeV blazars at Fermi-LAT suggest the existence of the intergalactic void magnetic fields, 
which might be the remnants of the primordial magnetic fields~\cite{Neronov:1900zz,Tavecchio:2010mk,Ando:2010rb,Dolag:2010ni,Essey:2010nd,Taylor:2011bn,Takahashi:2013lba,Finke:2015ona,Biteau:2018tmv}.
They can also contribute to solving the mysteries of the Universe such as the baryon asymmetry of the Universe~\cite{Giovannini:1997gp,Giovannini:1997eg,Bamba:2006km,Fujita:2016igl,Kamada:2016eeb,Kamada:2016cnb} or dark matter~\cite{Kamada:2017cpk}. 
If the primordial magnetic fields were generated in the early Universe at a very high temperature, then we expect that a stochastic GW background is produced through the photon-graviton conversion and serves as a smoking-gun of these scenarios.

In this paper, we examine the photon-graviton conversion process in the early Universe 
and evaluate the stochastic GW background today. 
We find that 
the process is most effective when the process onsets, {\it i.e.}, 
at the magnetic field generation or reheating,
since its efficiency at high energy scales is so strong that dilution due to the cosmic expansion
does not overwhelm it, 
and that  the stochastic GW background carries information of thermal photons in its spectrum. 
As a result, the information of the temperature or the time when the process starts is embedded in the location of the peak frequency.
For example, if the process starts at the Hubble parameter $H \simeq 6 \times 10^{13}$ GeV, 
the peak frequency comes to $f\simeq 10^2$ GHz. 
It also has a weak but nonvanishing dependence on the number of relativistic degrees of freedom. 
If we have a huge number of degrees of relativistic freedom in a thermal bath at the onset of the process, the peak shifts to lower frequencies, 
and hence it can also be a probe of the number of relativistic degrees of freedom. 
The peak height is determined by the strength of long-range magnetic fields
and we can give its theoretical upper bound by requiring the energy density of magnetic fields do not exceed that of thermal plasma. 
Unfortunately, we find that the GW amplitude cannot be large so that it can be detected by GW experiments in the foreseeable future.
But we believe our study 
enhances the motivation to improve the sensitivity of GW experiments.

The rest of the paper is organized as follows. 
In Sec.~\ref{sec:conversion}, the photon-graviton conversion process in the presence of long-range magnetic fields is reviewed. 
Then, we estimate the probability of this process in the thermal plasma in Sec.~\ref{sec:probability} and 
calculate the stochastic GW spectrum in Sec.~\ref{sec:spectrum}. 
Sec.~\ref{sec:conclusion} is devoted to conclusions and discussions.

\section{Photon-graviton conversion}\label{sec:conversion}

In this section, we review the core of the mechanism, that is, 
how gravitons are converted from photons by the 
background magnetic fields
(in the Minkowski spacetime)~\cite{Gertsenshtein1962,DeLogi:1977qe,Zeldovich:1983cr,Cillis:1996qy}.
To develop the intuitive understanding, we consider a simplified setup.
Namely, we ignore the effective mass of the photon in thermal plasma
and also neglect the back-reaction from the produced graviton to the photon,
while these effects are fully taken into account in the next section.

We here examine the conversion of photons into gravitons as waves (or EMWs into GWs)
in the presence of static magnetic fields. 
The convenient choice to investigate the evolution of the GWs is the transverse and traceless (TT) gauge, 
where the metric fluctuation $h_{\mu\nu}\equiv g_{\mu\nu} - \eta_{\mu\nu}$ satisfies $h^{0\mu}=0$, $h^i_i =0$,  and $\partial^i h_{ij}=0$. 
We will use $\eta_{\mu\nu}=(1,-1,-1,-1)$ notation.
Then the equation of motion for the GWs, $h_{ij}^\mathrm{TT}$, is obtained by linearizing the Einstein equation in the TT gauge as  
\begin{equation}
(\partial_t^2-\nabla^2)h_{ij}^\mathrm{TT} (t,\bm x)=\frac{2}{\Mpl^2}T_{ij}^\mathrm{TT},
\label{Classic EoM}
\end{equation}
where $T_{ij}^\mathrm{TT}$ is  the TT component of the energy momentum tensor of matter fields (including gauge fields)
and $\Mpl\approx 2.43\times 10^{18}\GeV$ is the reduced Planck mass.
The energy momentum tensor of the electromagnetic field is given by
\begin{equation}
T_{\mu\nu}^\mathrm{em}=\frac{1}{4}g_{\mu\nu}F^{\alpha\beta}F_{\alpha\beta}-F^{\ \alpha}_{\mu} F_{\nu\alpha},
\end{equation}
where
$F_{\mu\nu}\equiv \partial_\mu A_{\nu}-\partial_\nu A_{\mu}$ is the field strength of the photon field $A_\mu$.
Defining the electric and magnetic fields as 
\begin{equation}
E_i \equiv -F_{0i}, \quad B^i \equiv \frac{1}{2} \epsilon^{ijk} F_{jk} , \quad \text{or} \quad F_{ij} = \epsilon_{ijk} B^k, 
\end{equation}
with $\epsilon^{ijk} (\epsilon_{ijk})$ being the Levi-Civita symbol, 
$T_{ij}^\mathrm{em}$ is expressed as
\begin{equation}
\begin{split}
T_{ij}^\mathrm{em}&=\frac{1}{2} \delta_{ij} 
\delta^{kl} \left( E_k E_l+B_k B_l \right)
-E_i E_j - B_i B_j.\\[1ex]
\end{split}
\end{equation}
%
After going to the momentum space, the TT component of $T_{ij}^\mathrm{em}$
can be obtained by applying the projection tensor $\Lambda_{ij,kl}$, defined as
\begin{equation}
\begin{split}
\Lambda_{ij,kl}({\bm k}) &\equiv \left(P_{ik}({\bm k}) P_{jl}({\bm k}) - \frac{1}{2}P_{ij}({\bm k})P_{kl}({\bm k})\right), \\
P_{ij}({\bm k}) &\equiv \delta_{ij}-\frac{k_i k_j}{|{\bm k}|^2}, \label{projectionop}
\end{split}
\end{equation}
so that 
\begin{equation}
T^\mathrm{em, TT}_{ij} ({\bm k}) = \sum_{k,l} \Lambda_{ij,kl}({\bm k}) T^\mathrm{em}_{kl} ({\bm k}).  
\end{equation}

Now let us consider the photon-graviton conversion 
for an EMW that propagates along the $z$-axis
in the static and homogeneous background magnetic fields that run along the $x$-axis.
The gauge field configuration is given by
$A_\mu = {\bar A}_\mu +A^s_\mu$ where 
${\bar A}_\mu \equiv (0,0,0,B_0 y)$ is the background gauge field (magnetic field) and 
$A^s_\mu \equiv (0, (B_y/k) \sin (k (t-z)), -(B_x/k) \sin(k(t-z)),0)$ is the propagating gauge field (EWM).
We find
\begin{equation}
\begin{split}
B^i &= (B_0-B_x \cos (k(t-z)),  -B_y \cos (k(t-z)),0),  \\[1ex]
E_i &= (-B_y \cos(k(t-z)), B_x \cos(k(t-z)),0), 
\end{split}
\end{equation}
with which we obtain the energy momentum tensor for the electromagnetic field,
\begin{equation}
\begin{split}
&T_{ij}^\mathrm{em}({\bm x},t) \\
&=(B_0^2/2) \left(\begin{array}{cc} 
-1& 0 \\0&1
\end{array}\right) \\[1ex]
&\quad + \left(\begin{array}{cc}
B_0B_x \cos(k(t-z))&B_0B_y  \cos(k(t-z))\\
B_0B_y  \cos(k(t-z))& -B_0B_x \cos(k(t-z))
\end{array}\right) , \\[1ex]\label{emtreal}
\end{split}
\end{equation}
where $i,j$ runs 1 and 2. 
We now perform the Fourier transform and apply the projection operator defined in Eq.~\eqref{projectionop}.
Noting that for ${\bm p} = (0,0,p_z)$, the projection operator reads 
$P_{ij}({\bm p}) = \delta_{ij}$ for $i,j=1,2$ and 0 for others, 
we find the TT component of the energy momentum tensor in the momentum space as
\begin{equation}
\begin{split}
T_{ij}^\mathrm{em,TT}({\bm p},t) &=\delta(p_x) \delta (p_y) (\delta(p_z-k) e^{ikt} + \delta(p_z+k)e^{-ikt}) \\
&\quad\times (B_0/2) \left(\begin{array}{cc}
B_x & B_y  \\
B_y  & -B_x 
\end{array}\right) .
\end{split}
\end{equation}
Note that the non-propagating part in Eq.~\eqref{emtreal} does not contribute to the TT component. 
Going back to the real space, we obtain 
\begin{equation}
T_{ij}^\mathrm{em,TT}({\bm x},t)  = B_0 \cos(k(t-z))  \left(\begin{array}{cc}
B_x & B_y  \\
B_y  & -B_x 
\end{array}\right) ,
\end{equation}
for $i,j$ = 1,2 while the others vanish. 
By writing the TT component of the metric fluctuation in terms of $h_+(t,\bm x)$ and $h_\times(t,\bm x)$ as 
\begin{equation}
g_{\mu\nu}=\eta_{\mu\nu}+h_{\mu\nu}=\begin{pmatrix}1 & 0 & 0 & 0 \\
0 & -1+h_+ & h_\times & 0 \\
0 & h_\times & -1-h_+ & 0 \\
0 & 0 & 0 & -1 \\
\end{pmatrix},
\end{equation}
the linearized Einstein equation (Eq.~\eqref{Classic EoM}) reads
\begin{align}
(\partial_t^2-\partial^2_z)h_{+}(t,\bm x)=\frac{2}{\Mpl^2}B_0 B_x \cos[k(t-z)],
\\
(\partial_t^2-\partial^2_z)h_{\times}(t,\bm x)=\frac{2}{\Mpl^2}B_0 B_y \cos[k(t-z)].
\end{align}
Note that these equations are analogous to those of the driven (forced) oscillator.
The solution induced by the source term from the background magnetic fields is given by
\begin{equation}
h_+= (z+t)\frac{B_0 B_x}{2k \Mpl^2} \sin[k(t-z)].
\end{equation}
The solution for $h_\times$ can be obtained by replacing $B_x$ by $B_y$.
For instance, looking at $z=t+\pi/2k$, one can see the amplitude of 
the propagating GW grows in proportional to $t$,
as it is continuously produced by the source term.
It indicates that the GW is generated from the EMW in the magnetic field.

It should be noted that since we ignore the back-reaction from
the produced GWs to the EMW, formally the GWs would eventually acquire
infinite energy. In reality, however, once the amplitude of the GWs
become large enough, the back-reaction
becomes no longer negligible and the inverse process, namely, the conversion from the GW into the EMW,
should be significant. Consequently, the oscillation between the GW and the EMW takes place, as we see in the next section.

\section{The conversion in the universe}\label{sec:probability}

We now study the process of conversion from photons into gravitons in the thermal plasma of the early Universe. 
As long as the coherence length of background magnetic fields is sufficiently larger than the 
photon mean free path, the photon can be treated as an EMW in the static and homogeneous 
magnetic fields up to that length scale, and hence 
the basic idea in the previous section 
is applicable.
Since we will see that the time scale of the process is shorter than the Hubble time, we ignore the cosmic expansion and work with the Minkowski background.
On the other hand, here we take into account the non-negligible effects such as effective photon mass
and the inverse process, namely, the conversion from  gravitons into photons. 
As such, the photon and the graviton oscillates each other in a manner completely analogous to the neutrino oscillation or the axion-photon conversion~\cite{Raffelt:1987im,Sikivie:1983ip,Csaki:2001yk,Deffayet:2001pc}.
At the end of this section, we shall obtain the probability
of the conversion from a photon into a graviton, which allows us to examine the evolution of the graviton distribution function
in the next section.

Let us assume that stochastic magnetic fields with a sufficiently large coherence length exist 
in the radiation dominated Universe and electric fields are screened by the thermal plasma.
The linearized equations of motion for the graviton and photon that propagate along the $z$ direction in the background
large-scale magnetic fields are given by~\cite{Zeldovich:1983cr}
\begin{equation}
\begin{split}
\left(\partial_t^2-\partial_z^2\right)h_{ij}^\mathrm{TT}(t,z)&=\frac{2}{\Mpl^2}T_{ij}^\mathrm{em,TT},
\\[1ex]
\left(\partial_t^2-\partial_z^2+m_\gamma^2\right)A_i(t,z)
&=-\delta^{kj}\left(\partial_z h_{ij}^\mathrm{TT} \right)F_{kz}^{(\mathrm{bg})}, 
\end{split}
\end{equation}
for $i,j = 1,2$, where $F_{kj}^{(\mathrm{bg})} = \epsilon_{kjl} B^{(\mathrm{bg}) l}=-\delta^{lm}\epsilon_{kjl} B^{(\mathrm{bg})}_m$
is the field strength of the background large-scale magnetic fields and
the TT components of the energy momentum tensor are given by
\begin{equation}
\begin{split}
&T_{11}^\mathrm{em,TT} = - \,T_{22}^\mathrm{em,TT}
= B_x^{(\mathrm{bg})}  \partial_z A_y +B_y^{(\mathrm{bg})}  \partial_z A_x , \\[1ex]
&T_{12}^\mathrm{em,TT} = \, T_{21}^\mathrm{em,TT}
=-B_x^{(\mathrm{bg})}  \partial_z A_x +B_y^{(\mathrm{bg})}  \partial_z A_y .  \\[1ex]
\end{split}
\end{equation}
Here we work in the radiation gauge, $A_0 = 0, \delta^{ij} \partial_i A_j = 0$, and omit spatial derivatives on the background magnetic fields.
As mentioned in the above, we introduce the effective photon mass $m_\gamma$
emerged from the interactions between the photon and particles in the thermal plasma. Later we will also discuss the other effect coming from the photon interactions, namely, the photon mean free path.
By defining 
\begin{equation}
\begin{split}
A_+&=\frac{1}{\sqrt{B_x^{(\mathrm{bg})2}+B_y^{(\mathrm{bg})2}}}\left(B_y^{(\mathrm{bg})}A_x+B_x^{(\mathrm{bg})} A_y\right),  \\
A_\times&=-\frac{1}{\sqrt{B_x^{(\mathrm{bg})2}+B_y^{(\mathrm{bg})2}}}\left(B_x^{(\mathrm{bg})} A_x-B_y^{(\mathrm{bg})}A_y\right), 
\end{split}
\end{equation}
the EOMs are rewritten as 
\begin{equation}
\begin{split}
&\left( \partial_t^2 -\partial_z^2 + m_\gamma^2 \right) A_\lambda(t,z) \\
&\qquad \qquad \qquad +\sqrt{B_x^{(\mathrm{bg})2}+B_y^{(\mathrm{bg})2}}\, \partial_z h_\lambda(t,z) =0 , \\[1ex]
&\left( \partial_t^2 -\partial_z^2 \right) h_\lambda(t,z) \\
&\qquad \quad\,-2\Mpl^{-2}\sqrt{B_x^{(\mathrm{bg})2}+B_y^{(\mathrm{bg})2}}\, \partial_z A_\lambda(t,z) =0 ,
\end{split}
\end{equation}
where $\lambda = +, \times$.
%
%
%
It should be noted that the graviton and the photon are coupled to
the other only with the same polarization, $+$ or $\times$.

Assuming the three components of the background magnetic field have the same amplitude on average
$B_x^{(\mathrm{bg})2}=B_y^{(\mathrm{bg})2}=B_z^{(\mathrm{bg})2}\equiv B_T^2/3$,
we make the replacement $\sqrt{B_x^{(\mathrm{bg})2}+B_y^{(\mathrm{bg})2}} = \sqrt{2/3} B_T$.
We also redefine
the tensor field as ${\bar h}_\lambda \equiv (M_\mathrm{Pl}/\sqrt{2}) h_\lambda$. 
Then, focusing on a mode with a single frequency $\omega $,
\begin{equation}
A^\omega_\lambda(t,z)=\tilde{A}^\omega_\lambda(z) e^{-i\omega t},
\qquad
{\bar h^\omega}_\lambda(t,z)=\tilde{h}^\omega_\lambda(z) e^{-i\omega t},
\end{equation}
we obtain the set of EOMs as
\begin{equation}
\begin{split}
&\left( \omega^2+ \partial_z^2 - m_\gamma^2 \right) {\tilde A}_\lambda^\omega(z)
- (2/\sqrt{3}) (B_T/M_\mathrm{Pl}) \partial_z {\tilde h}_\lambda^\omega(z) =0, \\[2ex]
&\left( \omega^2+ \partial_z^2 \right) {\tilde h}_\lambda^\omega(z)
+ (2/\sqrt{3}) (B_T/M_\mathrm{Pl})  \partial_z {\tilde A}_\lambda^\omega(z) =0, \label{EOM}
\end{split}
\end{equation}
which indicate the mixing or conversion of the photon and the graviton. 
Note that we neglect the Hubble expansion and the decay of the overall amplitudes,
$|\dot{\tilde{A}}/\tilde{A}|, |\dot{\tilde{h}}/\tilde{h}| \sim H \ll \omega$.

We now examine the photon-graviton conversion rate with the EOMs of \eqref{EOM}. 
Since we are interested in the radiation dominated Universe with a high temperature $T>m_e$, 
the effective photon mass is dominated by the Debye mass, $m_\gamma^2 \simeq  m_D^2 \sim e^2 T^2$,
where $e=\sqrt{4\pi \alpha_e} \approx 0.3$ 
with $\alpha_e$ being the fine structure constant.
We also consider relatively large magnetic fields $B_T \lesssim T^2$. 
We further simplify the EOMs with the similar approximation taken in Ref.~\cite{Deffayet:2001pc}.
%
%
By assuming $\omega \gg m_\gamma, B_T/M_\mathrm{pl}$, we can approximate $-i\partial_z\simeq \omega$
while keeping $\omega + i \partial_z$.
%
%
Then, the EOMs are rewritten as
\begin{equation}
\begin{split}
\left[ \, \omega+i\partial_z + 
\mathcal{M} \,
\right]\begin{pmatrix}{\tilde A}^\omega_\lambda(z) \\
{\tilde h}^\omega_\lambda(z) \\
\end{pmatrix}=0,
\end{split}
\end{equation}
where
\begin{equation}
\begin{split}
\mathcal{M}\equiv
\begin{pmatrix} - \Delta_\gamma & -i \Delta_M \\
i \Delta_M & 0 \\
\end{pmatrix}
\equiv
\begin{pmatrix}-m_\gamma^2/2\omega & -iB_T/\sqrt{3}\Mpl \\
i B_T/\sqrt{3}\Mpl & 0 \\
\end{pmatrix}.
\end{split}
\end{equation}
%
%
%
%
%
These coupled equations can be solved in the same way as the neutrino oscillation.
Diagonalizing the mixing mass matrix $\mathcal{M}$ by a unitary matrix $U$, satisfying 
$U U^\dag = U^\dag U = \bm{1}$,
as
\begin{equation}
\begin{split}
U \mathcal{M} U^\dag &=
\begin{pmatrix}m_1 & 0 \\
0 & m_2 \\
\end{pmatrix}, \\[1ex]
m_{1,2} &=-  \frac{1}{2}\left[\Delta_\gamma
\pm \sqrt{\Delta_\gamma^2 + 4\Delta_M^2}\right],
\end{split}
\end{equation}
%
%
the EOMs for the rotated fields $\psi \equiv\begin{pmatrix}\psi_1 \\
\psi_2 \\
\end{pmatrix} \equiv U \begin{pmatrix}{\tilde A}^\omega_\lambda(z) \\
{\tilde h}^\omega_\lambda(z) \\
\end{pmatrix}$
are given by
$[\omega+i\partial_z+m_j]\psi_j=0$. 
Note that with the parameters we are interested in we have 
\begin{equation}
\Delta_\gamma \simeq T^2/\omega \gg T^2/M_\mathrm{Pl} \gtrsim B_T/\sqrt{3} M_\mathrm{Pl}  = \Delta_M ,
\end{equation}
as long as $\omega \ll M_\mathrm{Pl}$.  
Their solutions are easily obtained as
\begin{equation}
\psi_{j}(z)=e^{i(\omega+m_j)z} \psi_j(z=0),
\qquad (j=1,2). 
\end{equation}
Rewriting the solutions in terms of the original fields,
%
%
the photon-graviton conversion process can be seen from the 
solution for ${\tilde h}^\omega_\lambda(z)$ by setting ${\tilde h}^\omega_\lambda(0)=0$ 
so that
%
\begin{equation}
{\tilde h}^\omega_\lambda(z) = \frac{i}{2}e^{i\omega z}\sqrt{\frac{4\Delta_M^2}{\Delta^2_\gamma +4\Delta_M^2}}
(e^{im_2 z}-e^{im_1 z}){\tilde A}^\omega_\lambda(0).
\end{equation}
The probability that a photon traveling a distance $\Delta z = L$ is converted into a graviton
is then computed as
\begin{equation}
\begin{split}
P_{\gamma\to g}(L)&=|\langle {\tilde h}^\omega_\lambda (L)|{\tilde A}^\omega_\lambda(0)\rangle |^2
\\
&=\frac{4\Delta_M^2}{\Delta^2_\gamma+4\Delta_M^2}\sin^2
\left(\frac{\sqrt{\Delta_\gamma^2+4\Delta_M^2}}{2}L\right). \label{P full}
\end{split}
\end{equation}

Let us evaluate the typical length of the photon propagation $L$.  
The ensemble average of the photon-graviton conversion rate is given with that length. 
The collision of a photon can be understood as the collapse of the photon 
wave function and hence as a ``measurement''. 
Therefore, the photon-graviton conversion process can be regarded to be 
fixed for a wave-packet of the photon
with the propagation distance being the mean free path of the photon. 
It is evaluated as
$L\simeq \Gamma_\gamma^{-1}$,
where $\Gamma_\gamma$ is the rate of the photon scattering with charged particles in the thermal bath through the gauge interaction,
\begin{equation}
\Gamma_\gamma \simeq \alpha^2_e T.
\label{GG}
\end{equation}
Since we can approximate
%
%
%
%
%
%
%
$\sqrt{\Delta_\gamma^2+4\Delta_M^2} \simeq \Delta_\gamma$,  the argument of the sinusoidal function in Eq.~\eqref{P full} is evaluated as\footnote{Here we have omitted the momentum dependence of the mean free path
of the photon, but it will not change the result of Eq.~\eqref{arg} as far as we are interested in $\omega \lesssim T$.}
\begin{equation}
\Delta_\gamma L\simeq \frac{\Delta_\gamma}{\Gamma_\gamma}= \frac{2\pi T}{\alpha_e \omega}. \label{arg}
\end{equation}
Thus, as far as we are interested in $\omega\lesssim 10T$, we have 
$\Delta_\gamma L 
\gg 1$ and
the ensemble average of $P_{\gamma\to g}(L)$ can be obtained by replacing the sinusoidal function with $1/2$,
\begin{align}
\langle P_{\gamma \to g}(L)\rangle &\simeq 2\left(\frac{\Delta_M}{\Delta_\gamma}\right)^2
\simeq \frac{8\omega^2 B_T^2}{3e^4 T^4 \Mpl^2}
\notag\\[1ex]
&\approx 4 \times10^{-5}\left(\frac{\omega}{10^{14}\, \GeV}\right)^2\left(\frac{g_*}{100}\right)\Omega_B, \label{convrate}
\end{align}
where $\Omega_B$ is the energy fraction of the large scale magnetic field,
\begin{equation}
\Omega_B \equiv \frac{\rho_B}{\rho_{\rm tot}} =  \frac{15B_T^2}{\pi^2 g_* T^4}, 
\end{equation}
where $\rho_\mathrm{tot}=(\pi^2 g_*/30)T^4$ is the total energy density of the Universe
and $\rho_B=B_T^2/2$ is that of magnetic fields. 
Therefore, if the magnetic fields are produced at $T=10^{14} \, \GeV$, for instance,
with the energy fraction $\Omega_B$, 
$(0.004 \times \Omega_B) \%$ of photons at $\omega \sim T$ 
are converted into gravitons.


Note that our analytic estimate relies on the approximation $\partial_z \simeq i \omega$
but this is only valid for a sufficiently small $m_\gamma$.
However, since $m_\gamma \simeq 0.3 \, T$, it is not negligibly small for $\omega\lesssim T$. 
Thus our formula of Eq.~\eqref{convrate} does not have the accuracy beyond the order estimate.
Moreover, for a smaller $\omega < m_\gamma$, 
scattering processes between photons and other charged particles are no longer negligible and the picture discussed in the above breaks down. 
We expect that the photon-graviton conversion gets strongly suppressed 
and do not consider such a small momentum.

\section{The energy spectrum}\label{sec:spectrum}

Let us now evaluate the distribution function or the energy spectrum
of gravitons generated from the thermalized photons in the hot early Universe
through the mechanism discussed in the previous sections.
The generation of gravitons 
is described by the Boltzmann equation,
\begin{equation}
\left(\partial_t -H\omega\partial_\omega\right)f_{g}(t,\omega)
=\Gamma_{\gamma \to g} f_\gamma(t,\omega),
\label{Boltzmann1}
\end{equation}
where $H$ is the Hubble parameter, $f_g(t,\omega)$ is the distribution function of the graviton and that of the thermalized photon is given by the Bose-Einstein distribution,
\begin{equation}
f_\gamma=\frac{1}{e^{\omega/T}-1},
\end{equation}
with the temperature of the thermal bath $T$.
The conversion rate from the photon into
the graviton in a unit time $\Gamma_{\gamma \to g}$
can be evaluated with the same procedure in the case of the sterile neutrino production 
through oscillations \cite{Barbieri:1989ti,Barbieri:1990vx,Kainulainen:1990ds,Enqvist:1990ad,Enqvist:1991qj,Dodelson:1993je,Stodolsky:1986dx,Foot:1996qc,Abazajian:2001nj}. That is, 
we evaluate it in terms of the probability averaged over photons in the ensemble,
%
\begin{align}
\Gamma_{\gamma \to g}&=\frac{1}{2}\Gamma_{\gamma}\, \langle P_{\gamma \to g}(L)\rangle \simeq \frac{\omega^2 B_T^2}{12 \pi^2 M_\mathrm{Pl}^2 T^3},
\end{align}
where we have used Eq.~\eqref{convrate} and
$\Gamma_\gamma$ is the scattering rate (measurement rate) of the photon given in Eq.~\eqref{GG}.
This expression represents our evaluation 
that a photon is converted to a graviton with a rate 
$ \langle P_{\gamma \to g}(L)\rangle$ at a collision or a ``measurement'' 
that happens in a time scale $\Gamma_\gamma^{-1}$.
See Ref.~\cite{Stodolsky:1986dx} for the origin of the numerical factor 1/2.
We here also assume that the magnetic field coherence length is much larger than
the mean free path of the photon, which is usually the case.

Now we are ready to solve Eq.~\eqref{Boltzmann1}. Assuming that
the magnetic fields decay adiabatically after its generation at $T=T_i$
in the radiation dominated Universe\footnote{This is the case when the magnetic field coherence length is sufficiently large, which is consistent with our assumption. If it is short, magnetic fields evolve according to the magnetohydrodynamics and decay much faster. },
\begin{equation}
B_T= B_i \frac{T^2}{T_i^2}, \label{raddom}
\end{equation}
and changing the time variable from the cosmic time $t$ to the temperature $T$,
one can rewrite Eq.~\eqref{Boltzmann1} as
\begin{equation}
\left(T\partial_T+\omega\partial_\omega\right)f_g(T,\omega)
=-\mathcal{A} \frac{\omega^2}{TT_i}\frac{1}{e^{\omega/T}-1}, \label{Boltzmann2}
\end{equation}
with
\begin{equation}
\mathcal{A}\equiv \frac{B_i^2}{12\pi^2 H_i T_i \Mpl^2}
=\frac{H_i}{2\pi^2 T_i}\Omega_{Bi}, \label{aaa}
\end{equation}
where $H_i$ is the Hubble parameter at $T=T_i$ and
$\Omega_{Bi}\equiv B_i^2/(6\Mpl^2H_i^2)$ denotes the energy fraction of the magnetic fields at its generation.
This equation is analytically solved as
\begin{equation}
f_g(T,\omega)=\mathcal{A}\,\frac{\omega^2}{T^2T_i}\frac{T_i-T}{e^{\omega/T}-1}, 
\end{equation}
where we have set the boundary condition, $f_g(T_i, \omega)=0$.
Therefore at later times, the graviton distribution function is given by
\begin{equation}
f_g(T\ll T_i,\omega)=\mathcal{A} \frac{\omega^2/T^2}{e^{\omega/T}-1}.
\label{fg result}
\end{equation}
This result indicates that the photon-graviton conversion is most effective at $T\simeq T_i$ and 
for $T \ll T_i$ the graviton distribution function just redshifts.

The above equation holds
as long as Eq.~\eqref{raddom} is exact, that is, the entropy production does not occur 
so that
the thermal bath temperature evolves as $T \propto a^{-1}$ and $H \propto T^2$. 
This condition is violated when the number of relativistic particles
in the thermal equilibrium decreases.
Due to the entropy conservation in the matter sector, the temperature of the photon $T_\gamma$ 
depends not only the scale factor but also the number of relativistic degrees of freedom as
$T_\gamma \propto g_{*s}^{-1/3} a^{-1}$,
while the ``effective'' temperature of the graviton in Eq.~\eqref{fg result} is not affected, $T_g\propto a^{-1}$.  
The ratio between these temperatures at present is then given by
\begin{equation}
\frac{T_g(t_0)}{T_\gamma(t_0)}=\left(\frac{g_{*s}(t_0)}{g_{*s}(t_i)}\right)^{1/3}
\approx 0.33\left(\frac{g_{*s}(t_i)}{106.75}\right)^{-1/3}. 
\label{entropy production}
\end{equation}
Here $g_{*s}$ is the effective number of relativistic degrees of freedom for entropy,
and $t_0$ and $t_i$ denote the present time and the generation time
of the magnetic field, respectively.
We also used $g_{*s}(t_0)=43/11$.
Since the evolution equation for the gravitons holds 
by ignoring the right hand side and replacing $T$ with $T_g$ in Eq.~\eqref{Boltzmann2}, we can replace $T$ with $T_g$  to evaluate Eq.~\eqref{fg result} at a late time, as
\begin{equation}
f_g(T_g(t), \omega) = \mathcal{A} \frac{\omega^2/T_g^2(t)}{e^{\omega/T_g(t)}-1},
\end{equation}
which is our main result of the present paper. 

It is interesting to compare the peak frequencies of the intensities of the photon and the graviton, $I\propto \omega^3 f(\omega)$,
\begin{align}
&I_\gamma\propto\frac{\omega^3}{e^{\omega/T_\gamma}-1}
\quad\Longrightarrow\quad
2 \pi f_\gamma^\mathrm{peak} = \omega_\gamma^{\rm peak}\approx 2.82 T_\gamma,
\\
&I_g\propto\frac{\omega^5}{e^{\omega/T_g}-1}
\quad\Longrightarrow\quad
2 \pi f_g^\mathrm{peak}=\omega_g^{\rm peak}\approx 4.97 T_g.
\label{omegagpeak}
\end{align}
Therefore, by measuring the difference between the peak frequencies of the photon and the graviton,
one can probe the relativistic degrees of freedom at the time of primordial magnetogenesis through the following equation:
\begin{equation}
g_{*s}(t_i)\simeq21.3\left(
\frac{f_\gamma^{\rm peak}}{f_g^{\rm peak}}\right)^3.
\end{equation}

The gravitons investigated in the present study act as the stochastic GW background, 
which is often characterized by the differential energy fraction, $\Omega_\mathrm{GW} \equiv \frac{1}{\rho_{\rm tot}}\frac{\dd \rho_\mathrm{GW}}{\dd \ln \omega} $, where $\rho_\mathrm{GW} \equiv \int \ln \omega  (2 \omega^4) f_g(\omega) /2 \pi^2$. 
It is given by
\begin{equation}
\begin{split}
\Omega_{\rm GW}(\omega,t) 
&= \frac{\omega^4 f_g(T_g(t),\omega)}{\pi^2 \rho_{\rm tot}(t)}\\[1ex]
&= \frac{\Omega_{Bi} H_i}{2\pi^4 T_\gamma(t_i)}\frac{\omega^4}{\rho_{\tot}(t)}
\frac{\omega^2/T_g^2(t)}{e^{\omega/T_g(t)}-1}.
\end{split}
\end{equation}
Evaluating this equation at the present time $t_0$ with Eqs.~\eqref{entropy production} and \eqref{omegagpeak}, we obtain
\begin{equation}
\begin{split}
h^2\Omega_{\rm GW}(\omega,t_0) &\simeq 2.2 \times 10^{-11} \times \left(\frac{g_I}{106.75}\right)^{-\frac{13}{12}}\Omega_{Bi} \\
&\qquad \times \frac{(\omega/T_g (t_0))^6}{e^{\omega/T_g(t_0)}-1}
\left(\frac{H_i}{6 \times 10^{13}\GeV}\right)^{1/2},
\label{OmegaGW eq}
\end{split}
\end{equation}
with $g_I^{-13/12}\equiv (g_*(t_i))^{1/4} (g_{*s}(t_i))^{-4/3}$
and the Hubble constant $H_0=100h\,$km/s/Mpc. 
The peak angular frequency at the present is
\begin{align}
\omega_g^{\rm peak}&\simeq 4.97 T_g(t_0) \simeq 1.64T_\gamma(t_0)
\left(\frac{g_{*s}(t_i)}{106.75}\right)^{-1/3} \notag \\
\Rightarrow f_g^\mathrm{peak} &\equiv \frac{\omega_g^\mathrm{peak}}{2\pi}= 93 \mathrm{GHz} \left(\frac{g_{*s}(t_i)}{106.75}\right)^{-1/3}, \label{freqpg}
\end{align}
where we have used $T_\gamma(t_0)=2.73{\rm K}=3.57 \times 10^2$ GHz.
Note that the peak frequency depend on the cosmic temperature when the photon-graviton conversion occurred only through $g_i(t_i)$.
%
\begin{figure}[tbp]
  \begin{center}
  \includegraphics[width=80mm]{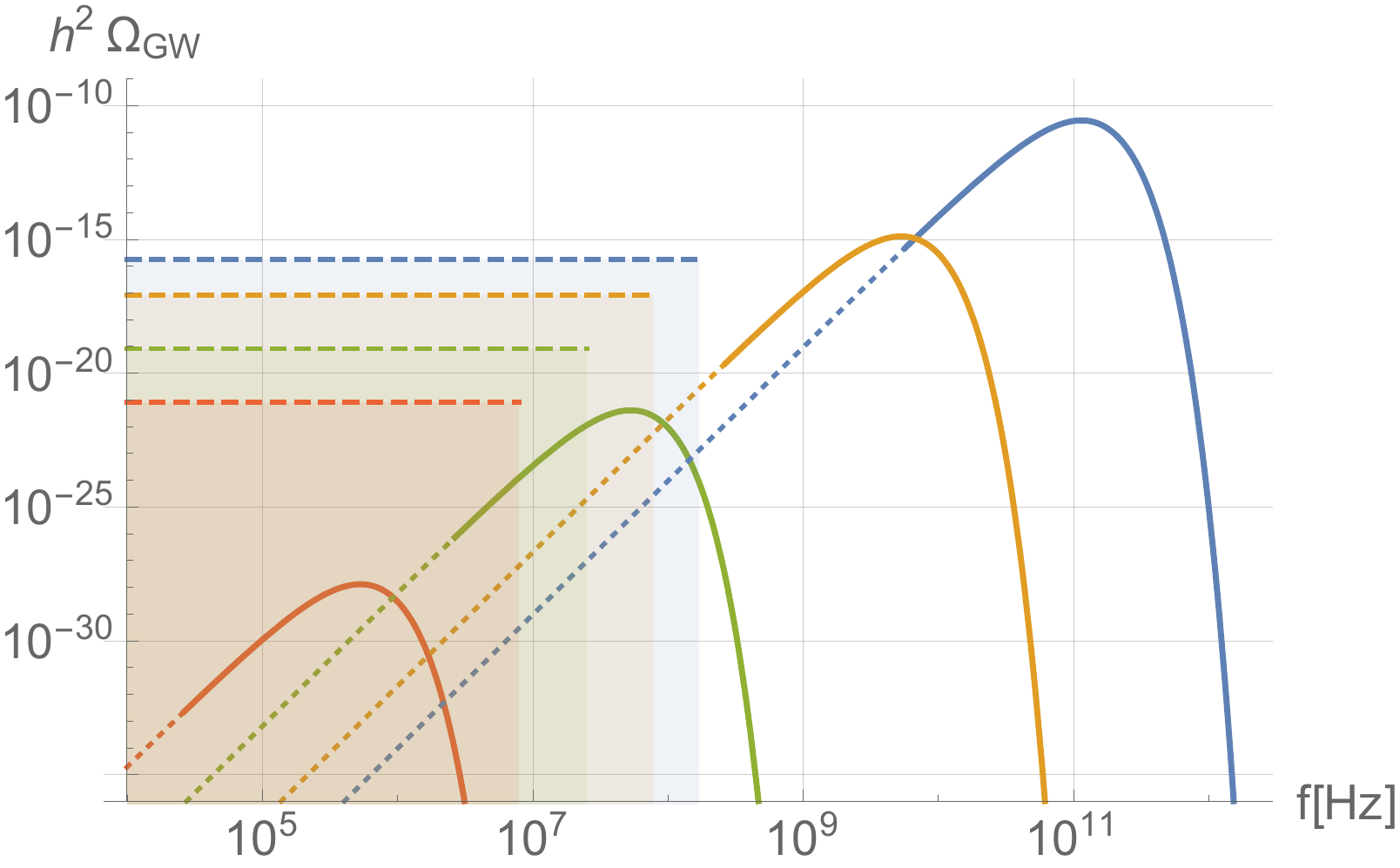}
  \end{center}
  \caption
 {$h^2\Omega_{\rm GW}$ derived in Eq.~\eqref{OmegaGW eq} is shown for $\Omega_{Bi}=10^{-2}$ and $H_i=H_\mathrm{end}=6\times 10^{13} \, \GeV$.
The effective number of relativistic degrees of freedom for entropy is taken as $g_I=10^2$ (blue), $10^6$ (orange), $10^{12}$ (green) and $10^{18}$ (red), respectively. 
$\Omega_{\rm GW}$ for $2\pi f<0.3 T_g$ are shown as dotted lines, because the approximation used in our analytic calculation is unreliable there. 
Dashed line represents the GW spectrum from inflation $h^2\Omega_\mathrm{GW}^{(\rm inf)}$ 
which overwhelms the GWs from photon-graviton conversion for $g_I\gg 10^{12}$.}
 \label{OmegaGW}
\end{figure}
%
In Fig.~\ref{OmegaGW}, we plot $h^2\Omega_{\rm GW}(\omega,t_0)$. 
As $g_I$ increases, both the peak frequency and the amplitude decrease.
Here we use the upper bound on the Hubble parameter $H_i\lesssim 6\times 10^{13}\GeV$, which comes from the non-detection of the CMB B-mode, as a reference value.
We also adopt $\Omega_{Bi}=10^{-2}$ to satisfy the constraint
from big bang nucleosynthesis $\Omega_{Bi}\lesssim 0.1$~\cite{Kawasaki:2012va}
and assume the comoving correlation scale of the magnetic field at the last scattering is shorter
than Mpc scale to evade the CMB bound~\cite{Ade:2015cva} (but is longer than the photon mean free path).
The constraint on the energy density of short-wavelength gravitational waves from 
the present additional relativistic degrees freedom
$h^2\int\ln\omega\, \Omega_{\rm GW}(\omega,t_0) < 1.2 \times 10^{-6}$~\cite{Pagano:2015hma}
is always satisfied.

Before concluding, let us comment on the comparison to 
other sources of GW background at high frequencies. 
One of the important and inevitable GW sources is  
the stochastic GW background from quantum fluctuations during inflation~\cite{Starobinsky:1979ty}.  It  
exhibits a scale-invariant spectrum up to the scale that corresponds to the Hubble scale at the end of 
 inflation. 
In the case of instant reheating, it is given as~\cite{Maggiore:1999vm,Guzzetti:2016mkm}
\begin{equation}
h^2\Omega_\mathrm{GW}^{(\rm inf)}  \simeq 1.7 \times 10^{-16}  \left(\frac{g_*^\mathrm{RH}}{106.75}\right)^{-1/3}\left(\frac{H_\mathrm{end}}{6 \times 10^{13}\mathrm{GeV}}\right)^2   \label{omegagwinf}
\end{equation}
for
\begin{equation}
f \leq f_\mathrm{RH}\equiv 0.17  \mathrm{GHz} \left(\frac{g_*^\mathrm{RH}}{106.75}\right)^{-1/12} \left(\frac{H_\mathrm{end}}{6 \times 10^{13}\mathrm{GeV}}\right), \label{freqend}
\end{equation}
where $H_\mathrm{end}$ is the Hubble parameter at the end of inflation 
and $g_{*}^\mathrm{RH}$ is the number of relativistic degrees of freedom at reheating. 
Note that at the edge of the spectrum, $f\simeq f_\mathrm{RH}$, another contribution from the gravitational particle production at the end of inflation, 
in which particles that include gravitons are generated due to the change of the cosmic expansion 
rate~\cite{Parker:1969au,Zeldovich:1971mw,Ford:1986sy}, also exists\footnote{The spectrum of this contribution depends on how the inflation ends and connects to the stage of the power-law expansion of the Universe~\cite{Hashiba:2018iff}},
but the amplitude of this contribution is as comparable as those from vacuum fluctuation (Eq.~\eqref{omegagwinf}). 
From these expressions (Eqs.~\eqref{freqpg} and \eqref{freqend}), we see that the frequency of these GW backgrounds is much smaller than 
that of the GWs from photon-graviton conversion for not so large $g_*$. 
However, for larger $g_*$,  
\begin{align}
g_{*}^\mathrm{RH} \simeq g_{*s}(t_i) > 9.6 \times 10^{12} \left(\frac{H_\mathrm{end}}{6 \times 10^{13} \mathrm{GeV}}\right)^{-4}, 
\end{align}
the frequency of the GW background from the photon-graviton conversion overlaps with those from 
inflationary fluctuations and the former can be hidden by  the latter, as we can see in Fig.~\ref{OmegaGW}. 
Note that this conclusion depends strongly on the reheating mechanism. 
If the reheating is followed by the long matter-like inflaton dominated era, 
the inflationary GW background are much diluted so that it might not hide the GWs from 
photon-graviton conversion. 
On the other hand, the inflationary contributions get larger 
for the kination scenario~\cite{Spokoiny:1993kt,Joyce:1996cp}, 
so that it is more likely to hide the GWs from photon-graviton conversion.

\section{Discussions}\label{sec:conclusion}

In this paper, we have explored the photon-graviton conversion process in the the presence of primordial magnetic fields 
and estimated the stochastic GW background which might be observed today. 
If the primordial magnetic fields survive until today, they can be the intergalactic magnetic fields which are reported to be detected by the blazar observations~\cite{Neronov:1900zz,Tavecchio:2010mk,Ando:2010rb,Dolag:2010ni,Essey:2010nd,Taylor:2011bn,Takahashi:2013lba,Finke:2015ona,Biteau:2018tmv}.
We found that the conversion process is the most effective at the magnetic field generation or reheating and that the information of the temperature and the number of relativistic degrees of freedom when the process starts is embedded in 
the peak frequency of the GW spectrum.
We gave the upper bound on the peak height
by requiring the energy density of magnetic fields do not exceed that of the thermal plasma. 
While the GW amplitude cannot be large enough to be detected anytime soon,
this study 
increases the motivation to improve the sensitivity of GW experiments. 
Indeed, the new proposals and attempts to detect very high frequency GWs have been made~\cite{Cruise:2006zt,Akutsu:2008qv,Cruise:2012zz,Ito:2019wcb,Ejlli:2019bqj}. In analogy with classic (electromagnetic-wave) astronomy, we expect that gravitational-wave astronomy will also develop technologies to observe GWs with a wide range of frequencies. In the future, therefore, the prediction of our study may be relevant and provide a novel way to explore the early universe.

It should be mentioned that a gauge field involved in the current process generating gravitons does not have to be the 
electromagnetic or hypercharge gauge field in the Standard Model (SM) of particle physics. 
If high energy hidden photons and their strong magnetic field counterpart exist in the early Universe, 
they can also generate a stochastic GW background. 
Hidden $U(1)$ gauge bosons have been explored intensively as a well-motivated possibility of
physics beyond the SM~\cite{Essig:2013lka},
since they are the key to unveil dark sectors and to understand the gauge structure of the SM.  
Hidden photons can be also a good candidate of dark matter or dark radiation. 
Moreover, recently, hidden magnetic fields have been gotten interest in the cosmological and astrophysical 
context such as the relationship to the indirect generation of intergalactic magnetic fields~\cite{Choi:2018dqr,Kamada:2018kyi}
and the spectral modulation of high-energy gamma-ray observations~\cite{Choi:2018mvk}. 
Therefore, it is worth studying the possible hidden photon-graviton conversion in the early Universe
and its observational signatures to explore dark sectors.

\section*{Acknowledgements}

We would like to thank Ruth Durrer, Soichiro Hashiba, and Sachiko Kuroyanagi
for useful discussions.
The work of TF was supported by JSPS KAKENHI 17J09103 and 18K13537.
The work of KK was supported by JSPS KAKENHI, Grant-in-Aid for Scientific Research JP19K03842 
and Grant-in-Aid  for Scientific Research on Innovative Areas 19H04610.
YN is grateful to KEK for their hospitality during his stay when the paper was completed.

\appendix



\end{document}